\documentclass[prl,twocolumn,floatfix]{revtex4-1}
\usepackage{amsmath} 

\usepackage{graphicx} 

\usepackage{verbatim} 

\usepackage{color} 

\usepackage{subfigure} 

\newcommand{\vb}[1]{\mathbf{#1}}

\begin{document}

\title{How Metallic are Small Sodium Clusters?} 
\author{J. Bowlan}
\author{A. Liang} 
\author{W. A. \surname{de Heer}} 
\affiliation{School of Physics, Georgia Institute of Technology, 837 State St.  
Atlanta, GA, 30332, USA} \date{October 26 2010, revised December 24 2010} 

\begin{abstract}

Cryogenic cluster beam experiments have provided crucial insights into the
evolution of the metallic state from the atom to the bulk. Surprisingly, one of
the most fundamental metallic properties, the ability of a metal to efficiently
screen electric fields, is still poorly understood in small clusters. Theory
has predicted that many small Na clusters are unable to screen charge
inhomogeneities and thus have permanent dipole moments. High precision electric
deflection experiments on cryogenically cooled Na$_N$ ($N<200$) clusters show
that the electric dipole moments are at least an order of magnitude smaller
than predicted, and are consistent with zero, as expected for a metal. The
polarizabilities of Na clusters also show metal sphere behavior, with fine size
oscillations caused by the shell structure.

\end{abstract}

\maketitle

By definition, a classical metal is a material which cannot support an internal
electric field. An electric field $\vb{E}_{\textrm{ext}}(r)$ applied to a metal 
object of arbitrary shape will cause the charge density to rearrange so that
$\vb{E}_{\textrm{int}}=0$. A caveat of this property is that a metallic object 
cannot have a permanent electric dipole moment (or any other moment), since this
implies that there is a non-vanishing internal electric field \cite{jackson}.
This property of metals applies on the macroscopic level, but it is not \em a
priori \rm obvious that it applies to extremely small objects such as metal
clusters. The effectiveness of the screening can be experimentally tested by
measuring the electric dipole moments and polarizabilities.

Early experimental and theoretical work on metal clusters focused on the static
dipole polarizability and demonstrated that alkali metal clusters could be
approximately treated as small metal spheres \cite{deheer_physics_RMP93}.  This 
led to the well-known jellium model which allowed a self-consistent description 
of the electronic shell structure of small clusters . The spherical jellium
model predicts that the polarizability of an alkali cluster is $\alpha(N)=(R 
+\delta(N))^3$, where $R = r_s N^{1/3}$ is the classical cluster radius, $r_s$ 
is the Wigner-Seitz radius, $N$ is the cluster size (in atoms). $\delta(N)$ is a 
quantum correction to the radius, often referred to as the “spillout factor” 
since it indicates that the electronic screening actually extends beyond the 
classical cluster radius. To first order, $\delta(N)$ is constant and comparable 
to the Lang-Kohn value for jellium surfaces \cite{lang1970theory}. In more 
sophisticated calculations, $\delta(N)$ varies with cluster size and shows 
non-trivial shell structure effects 
\cite{ekardt1986static,manninen_electronic_PRB86,
puska_electronic_PRB85}.

The spherical jellium model is clearly flawed: a small metal cluster is not
even approximately spherical \cite{deheer_physics_RMP93, 
schmidt_optical_EPJD99}, and the ionic structure has
been shown to have significant effects on the thermodynamic properties,
\cite{haberland_melting_PRL05, hock_premelting_PRL09} and photoelectron spectra
\cite{kostko_structure_PRL07}. Nevertheless, many physical properties, including 
the polarizabilities \cite{knight_polarizability_PRB85, 
tikhonov_measurement_PRA01, rayane_static_EPJD99} are surprisingly 
well-described. The existing experimental data on Na cluster polarizabilities 
only sparsely covers the range of cluster sizes, and the experiments were done 
at temperatures where the clusters are liquid \cite{haberland_melting_PRL05, 
hock_premelting_PRL09}.  However, as we show here, essential features of the 
jellium model are still observed even in high precision measurements, at 
cryogenic temperatures where the clusters are expected to be rigid with few or 
no excited vibrations (20 K).

Electric dipole moments are also expected to be highly sensitive to the 
electronic screening, and dipole moments have been observed in many
metal cluster systems (e.g. Nb, V, Ta \cite{moro_ferroelectricity_SCIENCE03, 
yin_electron_JSNM08}, and Sn and Pb 
\cite{schaefer_structure_JPCA08,schaefer_electric_JCP08}).  An asymmetric 
cluster without inversion symmetry is expected to have an electric dipole 
moment, and its magnitude depends on how effectively the charge inhomogeneity of 
the ion cores is screened by the valence electrons.  In the case of Pb clusters, 
the link between reduced screening and dipole moments is supported by a recent 
experiment\cite{senz2009core}  which found reduced core-hole screening in the 
same size range where dipole moments were observed 
\cite{schaefer_electric_JCP08} \footnote{Dipole moments were observed in Nb 
clusters (and 10 alloys) up to N = ~100.  A rigorous theoretical explanation 
that can account for the loss of screening and all of the related experimental 
phenomena is currently lacking.  See Ref.~\cite{yin_electron_JSNM08} for more 
details and references to theoretical work}.  For small clusters this reduced 
screening has been explained as a consequence of partial localization of the 
electrons due to change in the bonding character and low 
coordination\cite{senz2009core}. An all-electron quantum chemical calculation 
has predicted that similar effects will lead to dipole moments in Na clusters 
\cite{solovyov_structure_PRA02}.  Our experiment shows that the electric dipole 
moments are much smaller, and that metallic screening is not well described by 
theory even for a cluster as small as Na$_3$.  The failure of theory to 
correctly describe static screening in metal clusters is a serious outstanding 
problem.

Electric dipole moments and polarizabilites are ideally measured using
cryogenic molecular beam deflection methods. A beam of neutral metal clusters
is produced, deflected and detected using methods that have been previously
described \cite{deheer_physics_RMP93} (see 
Refs.~\cite{moro_ferroelectricity_SCIENCE03,yin_electron_JSNM08} for 
experimental details and parameters). Briefly, cryogenically cooled sodium 
clusters are produced in a laser vaporization (Nd:YAG 532 nm; 5 mJ/pulse) 
cluster source operating at 20 K. The beam velocity is measured with a 
mechanical chopper. The cluster beam is collimated (~$0.1$ mm slits) and passes 
through the pole faces of an inhomogeneous electric field ($E=85$ kV/cm, 
$dE/dz$=218 kV/cm$^2$). The clusters then deflect due to the force caused by the 
electric field gradient on the electric dipole that has an intrinsic component 
and an induced component. The induced dipole moment causes a uniform deflection 
of the cluster beam, while the intrinsic dipole moment (primarily) causes a 
broadening of the beam (see below for details). The cluster beam then enters a 
position sensitive time of flight mass spectrometer, that simultaneously 
measures the mass and deflection for all clusters in the beam. This method has 
been previously used to measure the electric dipole moments of large polar 
molecules and clusters \cite{broyer_structure_CRP02}.

\begin{figure}[h!]
\centering 
\subfigure{ 
\subfigure{\includegraphics[width=1.6in]{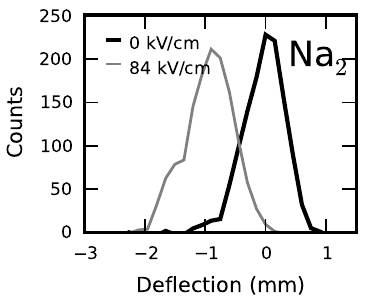}}
\subfigure{\includegraphics[width=1.6in]{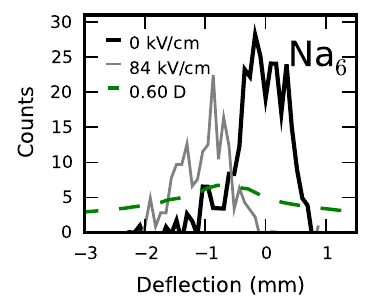}}}
\subfigure{
\subfigure{\includegraphics[width=1.6in]{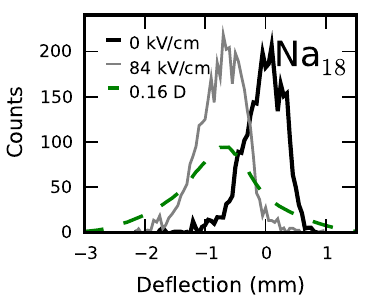}}
\subfigure{\includegraphics[width=1.6in]{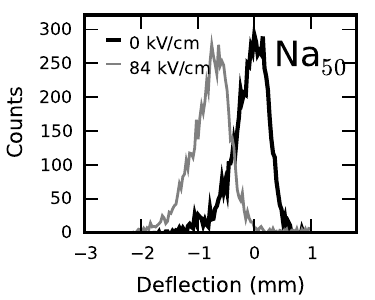}}}

\caption{Electric deflection profiles for Na$_{N}$ N=2,6,18, and 50.
The bold line shows the beam profile with the field off, the light line is with
the field on. The green dashed curve shows a simulation of the deflection
profile using the adiabatic rotor model \cite{bertsch1995magnetization}, with
the dipole moments calculated in Ref.~\cite{solovyov_structure_PRA02}. For
Na$_{6}$ a significant depletion of the beam intensity is observed which can be
explained by an isomer with a large dipole moment.}

 	\label{fig:na_profiles} 
\end{figure}

For every species in the beam, we measure a distribution of polarizations. We 
assume that the induced polarization $P=P_\alpha + P_p$ is due to
two effects: the electronic polarizability $P_\alpha = \alpha E$, and the dipole 
moment $p$ projected onto the field, time-averaged over the rotational motion 
$P_p = \langle p_z \rangle_t$. For a metal spheroid, $\alpha$ is an average of 
the principal polarizabilities\cite{tikhonov_measurement_PRA01}.  Because $P_p$ 
depends on the initial conditions (orientation, energy, and angular momentum) 
when a cluster adiabatically enters the deflector, the ensemble of clusters
shows a distribution of polarizations $\rho(P_p)$. Clusters will in general be 
deflected toward both the high and low field directions, depending on their 
initial orientation.  The observed deflection profile is thus a convolution of 
the beam profile with $\rho(P)$, and the signature of the dipole moment is a 
broadening of the molecular beam, which we measure by $\Delta \sigma = 
\sqrt{\sigma_{\textrm{on}}^2 - \sigma_{\textrm{off}}^2}$. (where 
$\sigma_{\textrm{on}/\textrm{off}}$ are the width of the peaks with the field on 
or off, respectively)

To derive a quantitative relation between the dipole moment $p$ and the beam
broadening $\Delta \sigma$, we use the adiabatic rotor model developed by
Bertsch and others \cite{bertsch1995magnetization,broyer_structure_CRP02}. This
model uses classical rigid-body mechanics to calculate $P_p = \langle p_z
\rangle_t$. For Na$_{10}$ at 20 K, the rotational constant $B = \hbar^2/2I
\approx 1 \mu$eV, so $2B/kT \approx 0.001$, thus the rotational levels are
effectively continuous, and classical mechanics applies.  For a spherical rotor in
the $pE/kT \ll 1$ limit, the model predicts a polarization distribution with the 
analytic form: $\rho(P) = (1/2p) \log |p/P|$ \cite{bertsch1995magnetization}.  
The variance of this distribution is $p^2/9$ so the deflection profile of a 
cluster with $p$ will show $\Delta \sigma = p/3$. For our experiment, $p=0.1$ D, 
gives $pE/kT \approx 0.01$, so the asymptotic regime $pE/kT \ll 1$ applies.  The 
structure of the cluster also effects the deflection profile. For symmetric tops 
($R_1\neq R_2=R_3$), the quantitative relation between $p$ and $\Delta \sigma$ 
is slightly different when $pE/kT \ll 1$.  Simulations for $R_1/R_3 = 1.4$ show 
that the relation $p=3 \Delta \sigma$ holds to within 7\%. Na clusters are known 
to show triaxial distortions, and there has been experimental and theoretical 
work \cite{elrahim_asymmetric_JPCA05} suggesting that a polar asymmetric rotors 
will tumble chaotically in the field if perturbed. This explanation was invoked 
to explain deflection experiments on biomolecules 
\cite{elrahim_asymmetric_JPCA05} with dipole moments of ~6 D, that showed 
reduced broadening.  In our laboratory, we have performed deflection experiments 
on weakly polar, highly asymmetric metal clusters (e.g. the planar Au$_9$ 
cluster (0.28 D) \cite{bowlan_au9}) and observed no evidence of chaotic 
tumbling. In this case, the beam is still symmetrically broadened just as in the 
symmetric top case, and the $p=3 \Delta \sigma$ estimate agrees with the value 
from multiple quantum chemical calculations. The model also assumes that any 
dipole moment is fixed in the cluster’s structure, and that the cluster is a 
rigid object.  At 20 K, the clusters are well below both the melting temperature 
and the range of temperatures where softening effects like premelting are known 
to occur \cite{haberland_melting_PRL05,hock_premelting_PRL09}.

\begin{figure}[h!]
\centering 
\includegraphics[width=3.5in]{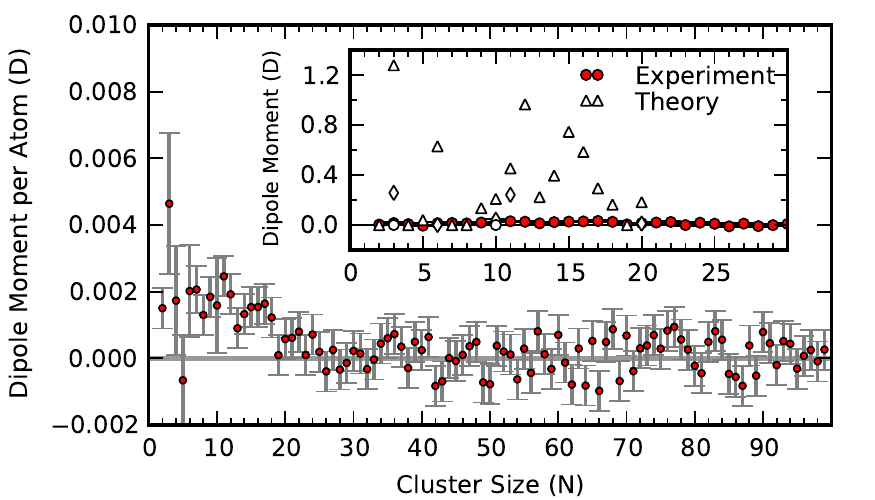}
\caption{Experimental dipole moments for Na clusters at 20 K, estimated from the 
spherical rotor model $p=3\sqrt{\sigma_{\textrm{on}}^2 - 
\sigma_{\textrm{off}}^2}$.  (inset) Comparison between the theoretical dipole 
moments calculated in Ref.~\cite{solovyov_structure_PRA02} and the experimental 
values above for small Na clusters.  Multiple theoretical values correspond to 
different isomers, dipole moments of zero (due to symmetry) were predicted for 
N=2,3,4,6,7,8,10, and 19.  Thus the discrepancy is greatest for N = 12-19.  The 
~1.2 D theoretical value for Na$_3$ is a high energy isomer that should not be 
present in our 20 K experiment.}
\label{fig:na_dipole}
\end{figure}

Per atom $p/N$ and total $p$ dipole moments estimated from the beam broadening
using $p=3\Delta \sigma$ are shown in Fig.~\ref{fig:na_dipole} Note that $p/N$ 
scatters around zero for all clusters $N>20$. For
$N<20$ there is a small amount of residual beam broadening which
cannot be explained away as an artifact.  \footnote{We have applied corrections
for other causes of broadening unrelated to a dipole moment: The dispersion in
the beam velocity is negligible. The inhomogeneity of the deflection
field causes a small but detectible broadening.  This correction accounts for 
all of the broadening observed in the atomic beams used for calibration (e.g. 
Li, Al, In)}.  For all of cluster sizes $p/N$ is less than 0.002 D per atom. 

The measured dipole moments appear to be greater than 0 for $N<20$ and Na$_3$
appears to have the largest $p/N$. Yet, its total moment is
only about 0.01 D. This measured value agrees with the measurements of Ernst
\cite{coudert_hyperfine_JCP02}, and is significantly lower than
Ref.~\cite{solovyov_structure_PRA02} which predicts a value of  ~0.3 D. It 
should be noted that there is agreement in the overall trend of the measured and 
calculated dipole moments, which indicates that the calculated shapes could be 
accurate but that screening is severely underestimated, even for very small 
clusters. Despite the success of simple shell models, a Na cluster is a 
many-body problem and practical calculations require approximations.  
Ref.~\cite{solovyov_structure_PRA02} uses a hybrid functional (B3LYP) to treat 
the exchange and correlation for the calculation of $\alpha$ and $p$.  This 
method is now known to show serious problems with bulk metallic systems 
\cite{paier2007does}.  There are many theoretical methods which deal with the 
many body problem at lower levels of approximation.  Understanding the origin of 
this error will require a comparison of these methods where the effect of the 
cluster structure is carefully controlled for.  Note that the electrostatic 
energy of a Na$_{20}$ cluster with a dipole moment of 0.1 D (far larger than 
what has been measured) is $E = \frac{p^2}{6 \epsilon_0 V} \approx 17\mu$eV.  
This suggests that correctly calculating the charge density requires high energy 
accuracy. 

Further note that Ref~\cite{solovyov_structure_PRA02} predicts two stable
isomers for Na$_6$. One is a planar triangle with $p=0$,
while the other is a pentagonal pyramid with $p\approx0.5$ D. Indeed, the 
observed intensity loss of Na$_6$ (Fig.~\ref{fig:na_profiles}) with applied 
field is consistent with two stable isomers, one which having a much larger $p$ 
than the other. Experiments are planned to further investigate Na$_6$. Note that 
for all other clusters there is no significant change in the total beam 
intensity when the electric field is turned on. 

\begin{figure}[h!] 
    \centering 
    \includegraphics{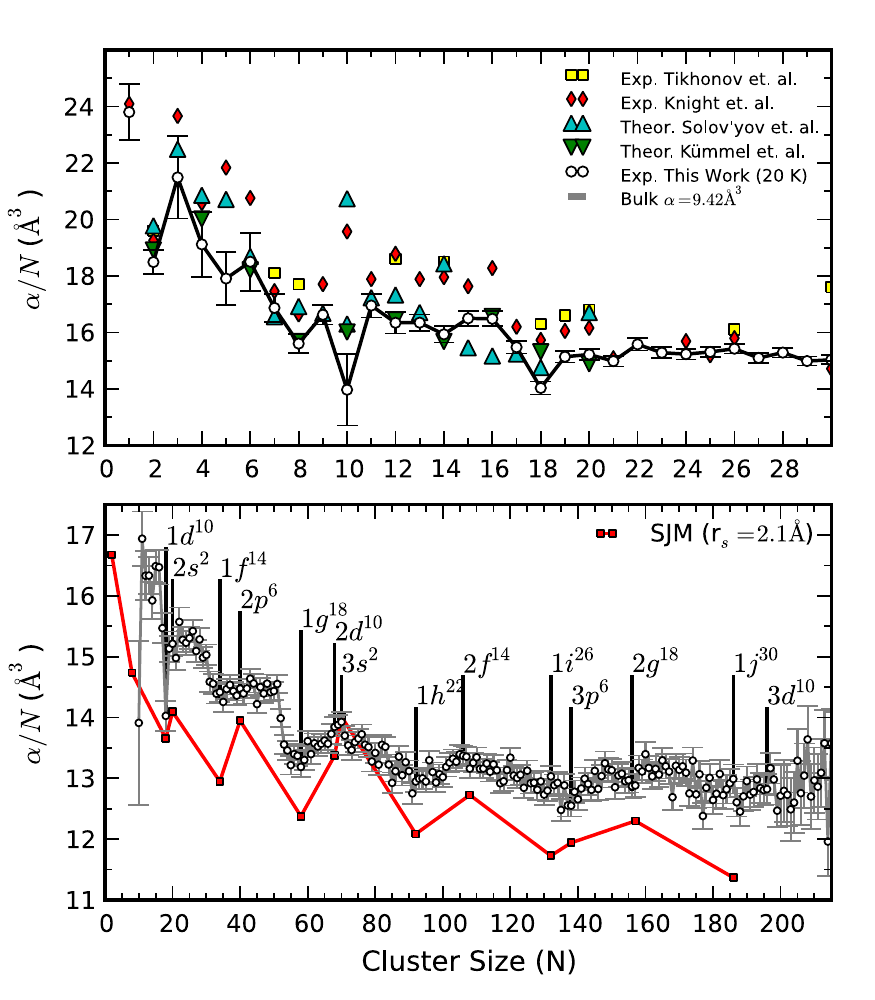}

    \caption{(upper) $\alpha/N$ of Na$_N$ ($N=1-30$) at a beam temperature of 20 
K.  Compared with previous higher temperature experiments 
\cite{knight_polarizability_PRB85,tikhonov_measurement_PRA01,rayane_static_EPJD99} 
$\alpha/N$ is systematically lower for most sizes and in better agreement with 
existing theory \cite{solovyov_structure_PRA02,kummel_polar_EPJD00}. $N=8$ and 
$N=18$ are minima as expected for a closed shell.  There is also a deep minimum 
at $N=10$, (although the variation between experimental runs is larger for 
$N=10$).  This supports the prolate structure for $N=10$ as predicted from the 
Clemenger-Nilsson model \cite{deheer_physics_RMP93}.
(lower) $\alpha/N$ for Na$_N$ ($N=10-200$) The shell closings have been marked.  
They coincide with the extrema of the oscillations about the descending trend.  
At $N=200$ the clusters are still far from the polarizability of bulk Na metal 
which is 9.4 \AA$^3/N$.  Shown for comparison is the prediction of the \em 
spherical \rm jellium model due to Ekardt \cite{ekardt1986static}}

\label{fig:na_alpha}
\end{figure}

We next turn to the high precision polarizability measurements for Na$_N$ $1
\leq N \leq 200$. (Fig.~\ref{fig:na_alpha}). First note that the measurements
of $\alpha(N)/N$ generally agree with previous reports. The overall decreasing
trend with increasing cluster size agrees with the simple approximation for the
polarizability of a conducting sphere with a spillout-enhanced radius
$\alpha(N) = (R + \delta)^3$.

Besides the overall decreasing trend, the present measurement also clearly
reveals variations in $\alpha/N$ with the shell structure, which were not
previously observed. (Fig.~\ref{fig:na_alpha}) Note that the minima in 
$\alpha/N$ correspond to spherical shell closings (e.g. $1p^6$ ($N=8$), 
$1d^{10}$ ($N=18$), $1f^{14}$ ($N=34$) , $1g^{18}$ ($N=58$) , $1h^{22}$ 
($N=92$), and within measurement error, $1h^{22}$ ($N=186$)). However shell 
closings do not always correspond to minima in $\alpha/N$. For example,
maxima are observed for $2d^{10}$ ($N=68$), $2f^{14}$ ($N=106$),
and perhaps $2g^{18}$ ($N=156$).  Hence, the systematic trend is, shell closings
with principal quantum number 1 are minimum in $\alpha/N$,
and those with principal quantum number 2, tend to be maxima.

These oscillations in the polarizability with the shell structure were predicted 
by Ekardt for Na \cite{ekardt1986static}, and by Puska and
co-workers for Li and Al clusters \cite{manninen_electronic_PRB86,
puska_electronic_PRB85} in the spherical jellium approximation.  Puska et. al.  
\cite{puska_electronic_PRB85} qualitatively explain this behavior as follows. By 
definition, for an electron in a quantum state with principal quantum number 1, 
there are no other electrons with the same angular momentum and lower principal 
quantum number. Consequently, electrons in these shells do not experience the 
Pauli repulsion from electrons in previously occupied shells with the same 
angular momentum. Therefore, the orbitals of these electrons penetrate deeper 
into the cluster and their spillout is reduced. In contrast, electrons in shells 
with principal quantum number 2 experience the Pauli repulsion from electrons 
with identical angular momentum in a previously filled shell (for example, 
electrons in the $2d$ shell are repelled by electrons in the $1d$ shell.) This 
repulsion enhances the spillout and causes $\alpha/N$ to increase as this shell 
is filled.

A triaxial distortion can also enhance the axis averaged $\alpha$ of a
cluster.  However, estimates of the magnitude of this effect using values of
the distortion parameter from photoabsorption experiments
\cite{schmidt_optical_EPJD99} shows that it is too small to account for the
magnitude of the oscillations. It is also noteworthy that, a significant
anomaly in the generally smooth trend is observed at $N=55$. This sudden drop
in $\alpha/N$ immediately before the electronic shell closing at $N=58$, and is 
likely caused by the geometric shell closing.

Overall the polarizabilities are systematically smaller than reported in 
previous experiments \cite{knight_polarizability_PRB85, 
tikhonov_measurement_PRA01, rayane_static_EPJD99}, and are in closer agreement 
with existing theory \cite{kummel_polar_EPJD00,solovyov_structure_PRA02}.
This effect has been predicted \cite{blundell_temperature_PRL00} and is related 
to thermal expansion.  It is surprising that theory gives $\alpha$ to within 
5-10\%, while the dipole moments are off by orders of magnitude, but this was 
already nearly the case with the spherical jellium model which has zero dipole 
moment by symmetry \cite{ekardt1986static}. 

In conclusion, the electric deflection measurement discussed here gives a
comprehensive picture of the response of small sodium clusters to static
electric fields. The nearly vanishing electric dipole moments, even for
clusters as small as the sodium trimer, demonstrates that the electric fields
surrounding alkali clusters are very small, as expected for a
classical metallic object. The observed dipole moments are much smaller than
predicted by quantum chemical methods, indicating a fundamental challenge for
the theoretical treatment of dipole moments in metallic clusters.


We gratefully acknowledge helpful discussions with Xiaoshan Xu. Funding was 
provided by the NSF under fund No.  R7418-DMR-0605894

\bibliography{na_metal}

\end{document}